\documentclass[11pt]{article} 

\usepackage{deauthor,times,graphicx} 
\usepackage{tcolorbox}
\usepackage{xpatch}
\usepackage{xspace}   

\newcommand{\oursys}{\textit{iDataScience}\xspace}
\newcommand{\hi}[1]{\vspace{.25em} \noindent {\bf #1}\xspace}
\newcommand{\llm}{\textsc{LLM}\xspace}
\newcommand{\llms}{\textsc{LLMs}\xspace}
\newcommand{\mllm}{\textsc{MLLM}\xspace}
\newcommand{\mllms}{\textsc{MLLMs}\xspace}

\makeatletter

\makeatletter

\begin{document}

\title{Data Agent: A Holistic Architecture for Orchestrating Data+AI Ecosystems}
\author{Zhaoyan Sun, Jiayi Wang, Xinyang Zhao,  Jiachi Wang, Guoliang Li\\ \small Department of Computer Science, Tsinghua University, Beijing, China\\  \small \{liguoliang@tsinghua.edu.cn\}}

\maketitle

\begin{abstract}
Traditional Data+AI systems utilize data-driven techniques to optimize performance, but they rely heavily on human experts to orchestrate system pipelines, enabling them to adapt to changes in data, queries, tasks, and environments. For instance, while there are numerous data science tools available, developing a pipeline planning system to coordinate these tools remains challenging. This difficulty arises because existing Data+AI systems have limited capabilities in semantic understanding, reasoning, and planning. Fortunately, we have witnessed the success of large language models (LLMs) in enhancing semantic understanding, reasoning, and planning abilities. It is crucial to incorporate LLM techniques to revolutionize data systems for orchestrating Data+AI applications effectively.

To achieve this, we propose the concept of a `Data Agent' -- a comprehensive architecture designed to orchestrate Data+AI ecosystems, which focuses on tackling data-related tasks by integrating knowledge comprehension, reasoning, and planning capabilities. We delve into the challenges involved in designing data agents, such as understanding data/queries/environments/tools, orchestrating pipelines/workflows, optimizing and executing pipelines, and fostering pipeline self-reflection. Furthermore, we present examples of data agent systems, including a data science agent, data analytics agents (such as unstructured data analytics agent, semantic structured data analytics agent, data lake analytics agent, and multi-modal data analytics agent), and a database administrator (DBA) agent. We also outline several open challenges associated with designing data agent systems.
\end{abstract}
\section{Introduction}
\label{sec:intro}

In the past decade, the database community has made significant contributions to the Data+AI field~\cite{li2021ai}. On one hand, for AI4Data, our community leverages AI techniques to tackle offline NP-hard problems (e.g., index advisor~\cite{wu2024automatic,zhou2024breaking}, view advisor~\cite{han2023dynamic}, partition advisor~\cite{zhou2023grep}, knob advisor~\cite{zhao2023automatic}, hint advisor~\cite{marcus2021bao}), online NP-hard problems (e.g., query rewrite~\cite{zhou2021mcts}, plan enumeration~\cite{yu2022hybrid}), regression problems (e.g., cardinality estimation~\cite{wang2021face,sun2021exploration}, cost estimation~\cite{hilprecht2022zeroshot}, latency estimation~\cite{wehrstein2025foundation}), prediction challenges (e.g., workload prediction~\cite{zhou2020predict}), and data structure design issues (e.g., learned indexes~\cite{sun2023index}). These efforts primarily focus on machine learning model design and selection, as well as the design and selection of data/query/environment features. However, adapting these techniques to changes in data, queries, tasks, and environments poses a significant challenge, as they rely heavily on experts tuning to accommodate different scenarios. On the other hand, for Data4AI, our community extends database optimization techniques to ease the deployment of AI, including in-database machine learning (ML) training and inference~\cite{guo2025indatabase}, data preparation~\cite{chai2023goodcore}, data cleaning~\cite{siddiqi2023saga}, data integration~\cite{wang2025aop}, feature management~\cite{zhou2023febench}, and model management~\cite{moritz2018ray}. The main obstacle with these methods is achieving autonomous orchestration of system pipelines without labor-intensive involvement.

\begin{figure}[!t]
\centering
\includegraphics[width=.9\linewidth]{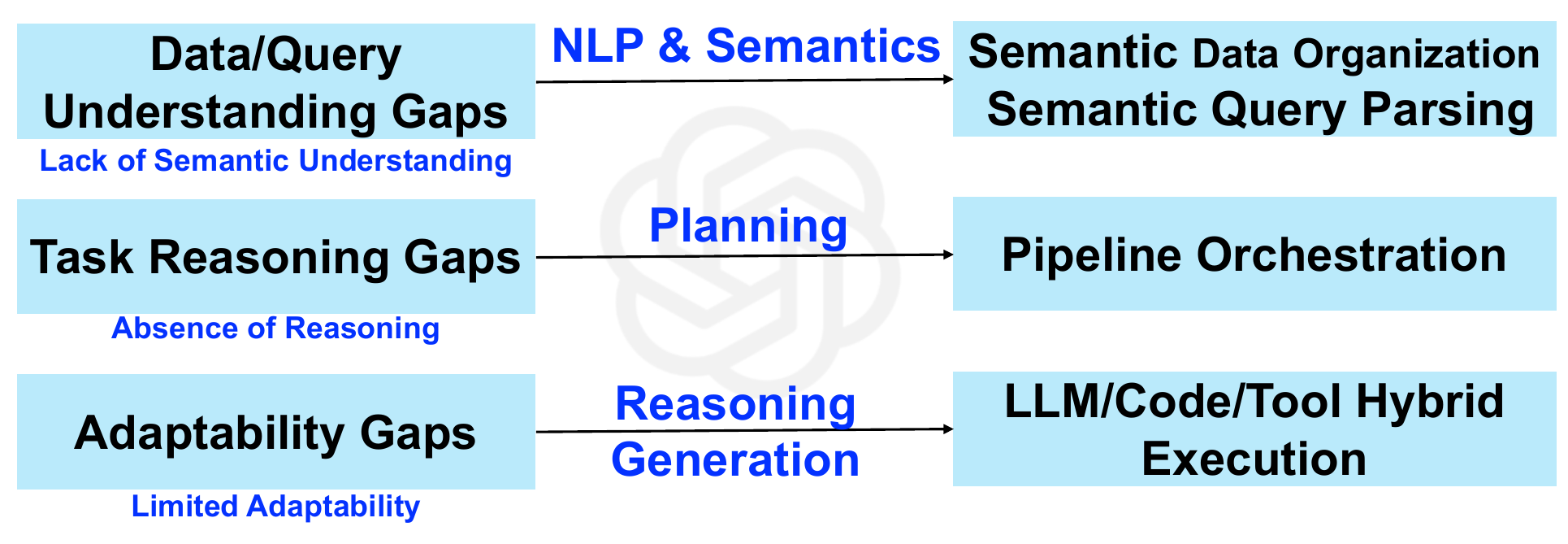}
\caption{Challenges of Data+AI Systems. \label{fig:challenges}}
\end{figure}

The core obstacle preventing existing Data+AI techniques from adapting to varying scenarios is their limited ability in semantic understanding, reasoning, and planning, as shown in Figure~\ref{fig:challenges}. Fortunately, large language models (LLMs) possess these capabilities~\cite{openai2024gpt4o,guo2025deepseek}, and we aim to leverage them to revolutionize Data+AI systems. To accomplish this, we propose the `Data Agent,' a comprehensive architecture designed to orchestrate Data+AI ecosystems by focusing on data-related tasks through knowledge comprehension, reasoning, and planning abilities. We outline several challenges in designing data agents:

Challenge 1: How can we understand queries, data, agents, and tools?

Challenge 2: How can we orchestrate effective and efficient pipelines to bridge the gaps between user requirements and the underlying heterogeneous data (e.g., data lakes)?

Challenge 3: How can we schedule and coordinate agents and tools to improve the effectiveness?

Challenge 4: How can we optimize and execute pipelines to improve the efficiency?

Challenge 5: How can we continuously improve pipeline quality with self-reflection?

We begin by proposing a holistic architecture to address these challenges. We present a robust architecture for developing data agents, which includes components for data comprehension and exploration, understanding and scheduling within the data engine, and orchestrating processes through pipeline management. Subsequently, we demonstrate several data agent systems, such as a data science agent, data analytics agents (including unstructured data analytics agents~\cite{wang2025aop}, semantic structured data agents, and data lake agents~\cite{iDataLake}), and a database administrator (DBA) agent~\cite{zhou2024dbot,sun2025dbot}. Finally, we identify some open challenges associated with designing data agent systems.

Our contributions can be summarized as follows:

(1) We introduce the data agent, which autonomously handles data-related tasks with capabilities for knowledge comprehension, automatic planning, and self-reflection.

(2) We present a holistic architecture for orchestrating a data agent system.

(3) We showcase three types of data agent systems: the data science agent, the data analytics agents, and the DBA agent.

(4) We provide several open research challenges related to data agents.

\section{Data Agent}
\label{sec:DA}

The Data Agent is designed to autonomously carry out data-related tasks with capabilities for knowledge comprehension, automatic planning, and self-reflection.

\begin{figure}[!t]
\centering
\includegraphics[width=.9\linewidth]{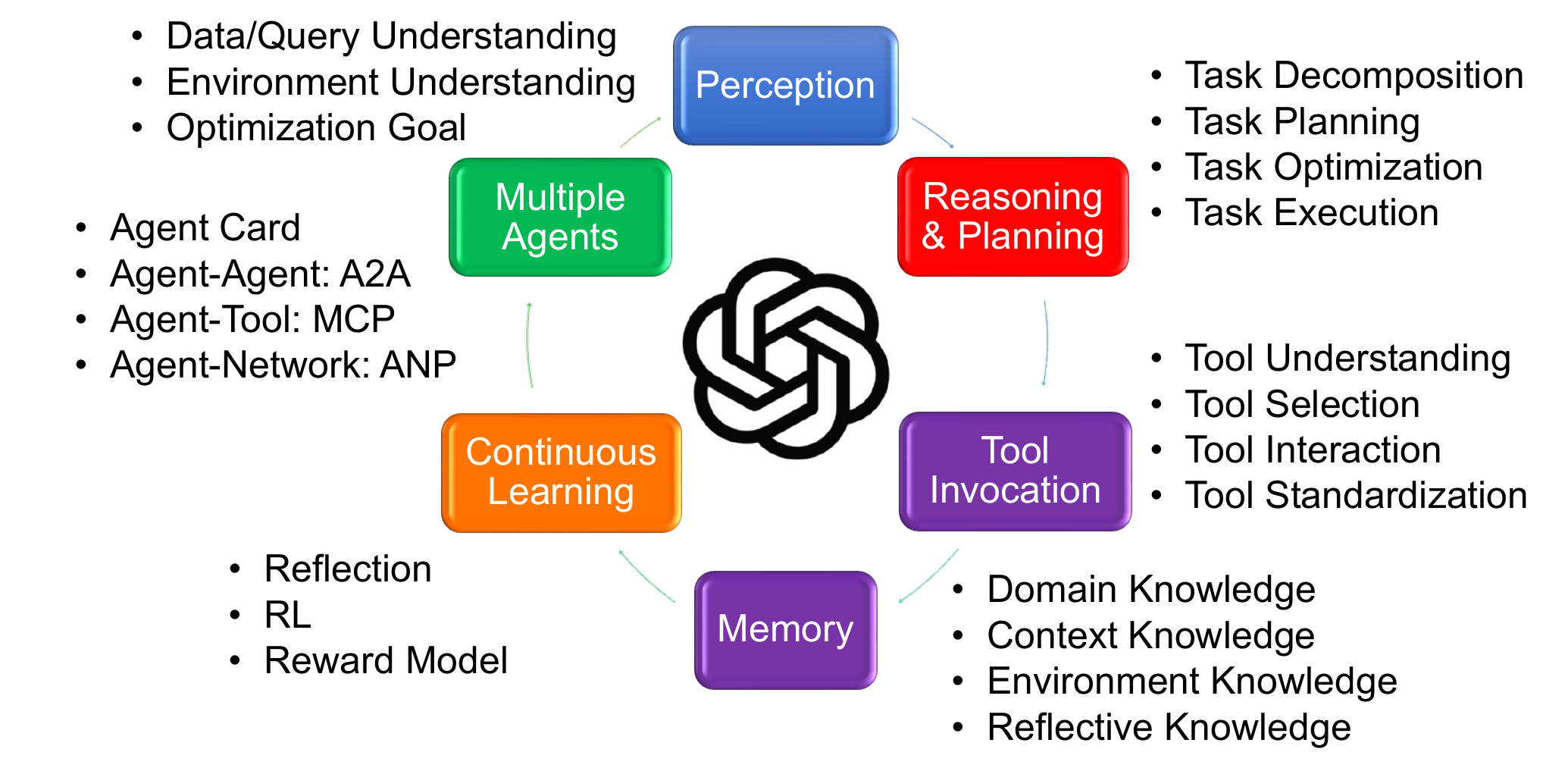}
\caption{Key Factors of Data Agents. \label{fig:challenges2}}
\end{figure}

Data Agents require to consider six key factors as shown in Figure~\ref{fig:challenges2}.

\hi{Perception:} This involves understanding the environment, data, tasks, agents, and tools. It requires aligning the Data Agent through offline fine-tuning or by preparing offline prompt templates.

\hi{Reasoning and Planning:} Planning focuses on creating multi-step pipeline orchestration, while reasoning involves making single-step decisions or actions. Each action may require further exploration of reasoning/planning or invoking a tool (to acquire domain data or knowledge).

\hi{Tool Invocation:} The agent can call upon tools to perform calculations, access domain-specific data, or provide instructions to environments. The Model Context Protocol (MCP) facilitates alignment between agents and tools, ensuring that information and states are exchanged in a standard format to prevent information drift~\cite{anthropic2024mcp}. Intermediate inference results from different models can be understood and reused across the system.

\hi{Memory:} This includes long-term memory, such as domain-specific and environmental knowledge, and short-term memory, like user context. Typically, a vector database is used to store and query these memory data. Other types of memory, such as reflective memory, will also be used to enhance planning abilities and performance.

\hi{Continuous Learning:} Continuously improving the agent to make it smarter is vital. This relies on self-reflection, reinforcement learning, and reward model techniques for self-improvement.

\hi{Multiple Agents:} Individual agents may struggle to handle diverse tasks effectively, as each agent has its own strengths but also limitations. Thus, integrating multiple agents to collaborate and coordinate is necessary for complex tasks. This approach enhances system robustness and improves parallelism and efficiency.

\begin{figure}[!t]
\centering
\includegraphics[width=.9\linewidth]{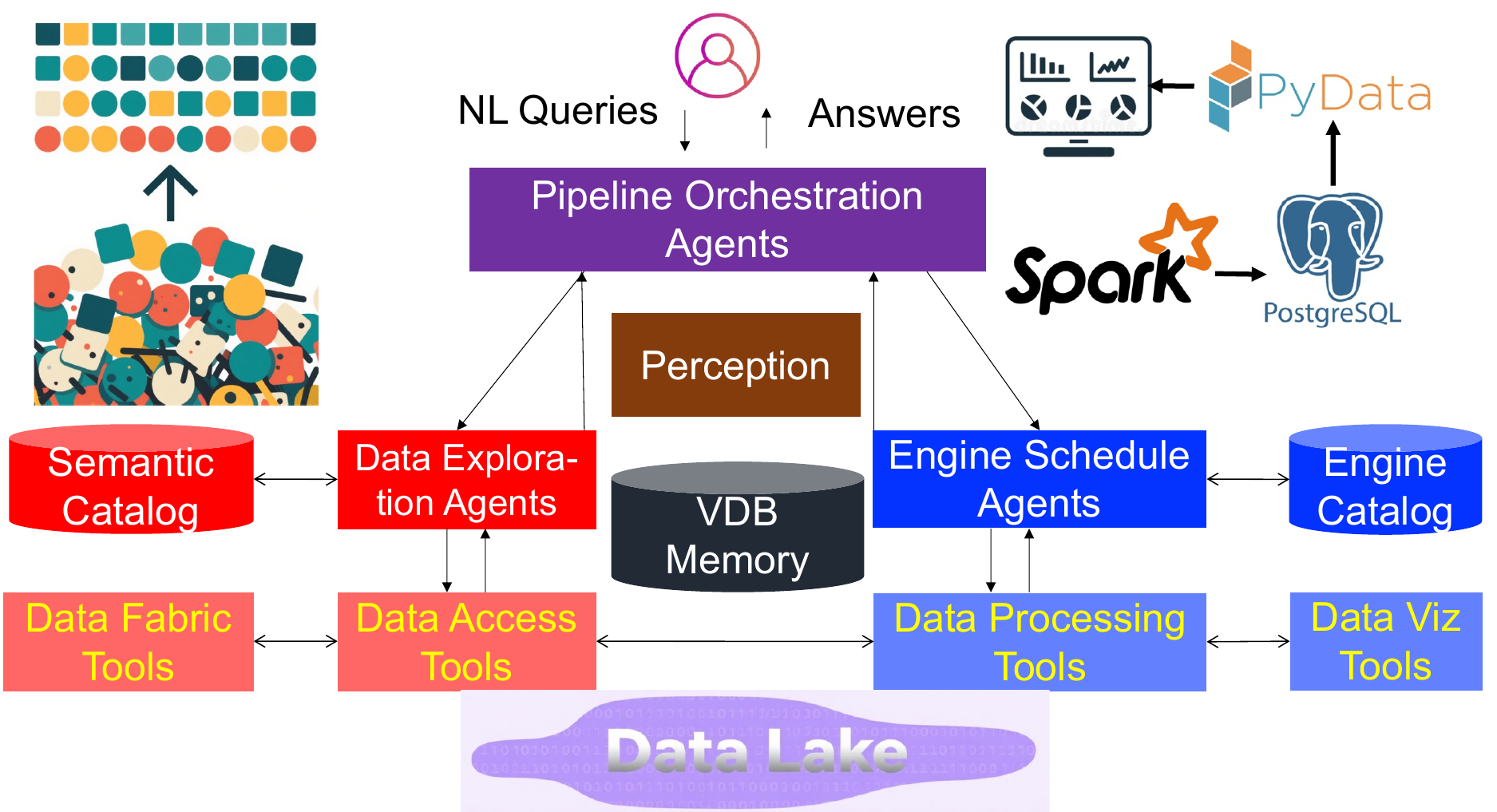}
\caption{Data Agents Framework. \label{fig:framework1}}
\end{figure}

\begin{figure}[!t]
\centering
\includegraphics[width=.9\linewidth]{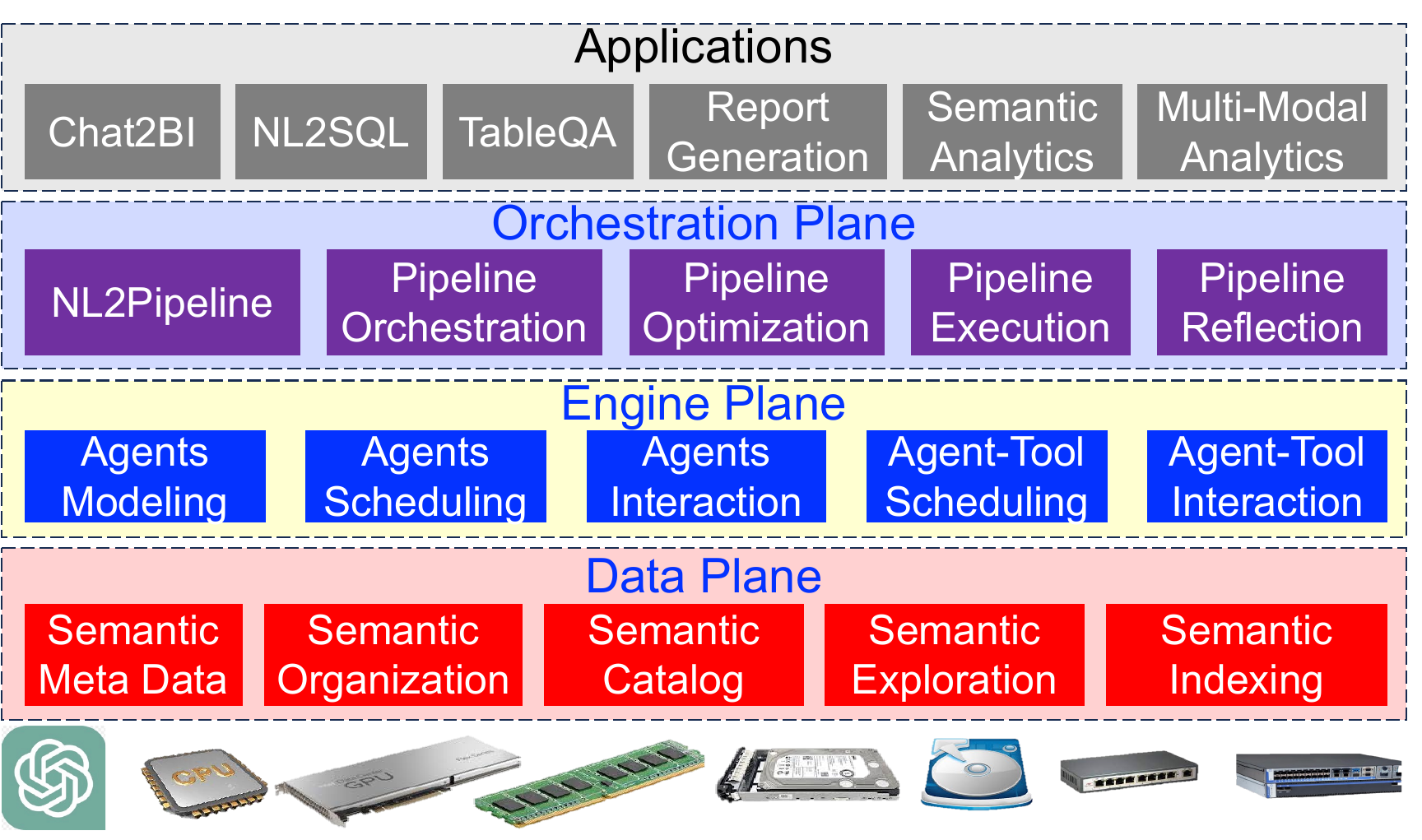}
\caption{Data Agents Architecture. \label{fig:framework2}}
\end{figure}

We propose a comprehensive architecture for building data agents, encompassing data understanding and exploration, data engine understanding and scheduling, and pipeline orchestration, as shown in Figure~\ref{fig:framework1}. Figure~\ref{fig:framework2} shows the detailed architecture.

\hi{Data Understanding and Exploration Agents} aim to organize and understand data to facilitate discovery and access by the agent. A unified semantic catalog offers a well-structured metadata system (e.g., schema and metadata index), enhancing data access performance. The data fabric provides a unified view of heterogeneous data by linking and integrating diverse data, allowing easy data retrieval by the agent. Semantic data organization and semantic indexes are also very improve to improve the data agent efficiency. Importantly, there are numerous tools for data preparation, cleaning, and integration. This component will also devise effective strategies to utilize these tools efficiently.

\hi{Data Engine Understanding and Scheduling Agents} focus on comprehending and scheduling data processing engines, such as Spark, DBMSs, Pandas, and PyData. Given that different agents and tools have varying skill sets, it is essential to profile the specific capabilities of each engine and coordinate them to execute complex tasks effectively. 

\hi{Pipeline Orchestration Agents} are responsible for generating pipelines based on user-input natural language (NL) queries and the data catalog.  They  break down complex tasks into smaller, manageable sub-tasks that can be executed sequentially or in parallel to achieve the overall goal. Given that both NL queries and the underlying data exist in an open-world context, these agents must leverage the understanding, reasoning, and self-reflection capabilities of large language models (LLMs) to create high-quality plans. Subsequently, the pipelines need to be optimized to improve latency, cost, or accuracy, and engine agents are invoked to efficiently execute the pipelines.

\hi{Memory} encompasses long-term memory, such as domain and environmental knowledge, as well as short-term memory, like user context and reflective context. Vector databases are typically used to manage this memory for enhancing the performance.

\hi{Perception} is tasked with comprehensively understanding the surrounding environments and the specific tasks at hand.

\hi{Agent-Agent Interaction}  is designed to coordinate and collaborate multiple agents to tackle decomposed sub-tasks. It comprises three key components: agent profiling and selection, agent interaction and coordination, and agent execution.
Agent Profiling and Selection involves building profiles for agents, enabling the system to choose the most suitable agents for specific tasks.
Agent Interaction and Coordination focuses on the coordination and interaction of multiple agents to effectively address sub-tasks. Thanks to agent-to-agent (A2A) protocols, we can facilitate communication between agents and synchronize their statuses via A2A~\cite{google2025a2a}.
Agent Execution aims to execute multiple agents either in a pipeline or in parallel, enhancing the system's fault tolerance and fast recovery.

\hi{Agent-Tool Invoking}  is utilized for calling upon appropriate tools. Given the multitude of data processing tools available, such as Pandas and PyData, it is necessary to select the right tool for each task. The challenge lies in profiling and scheduling these tools effectively. Fortunately, the Model Context Protocol (MCP) allows us to easily integrate new tools into the data agent system. 

\begin{figure}[!t]
\centering
\includegraphics[width=\linewidth]{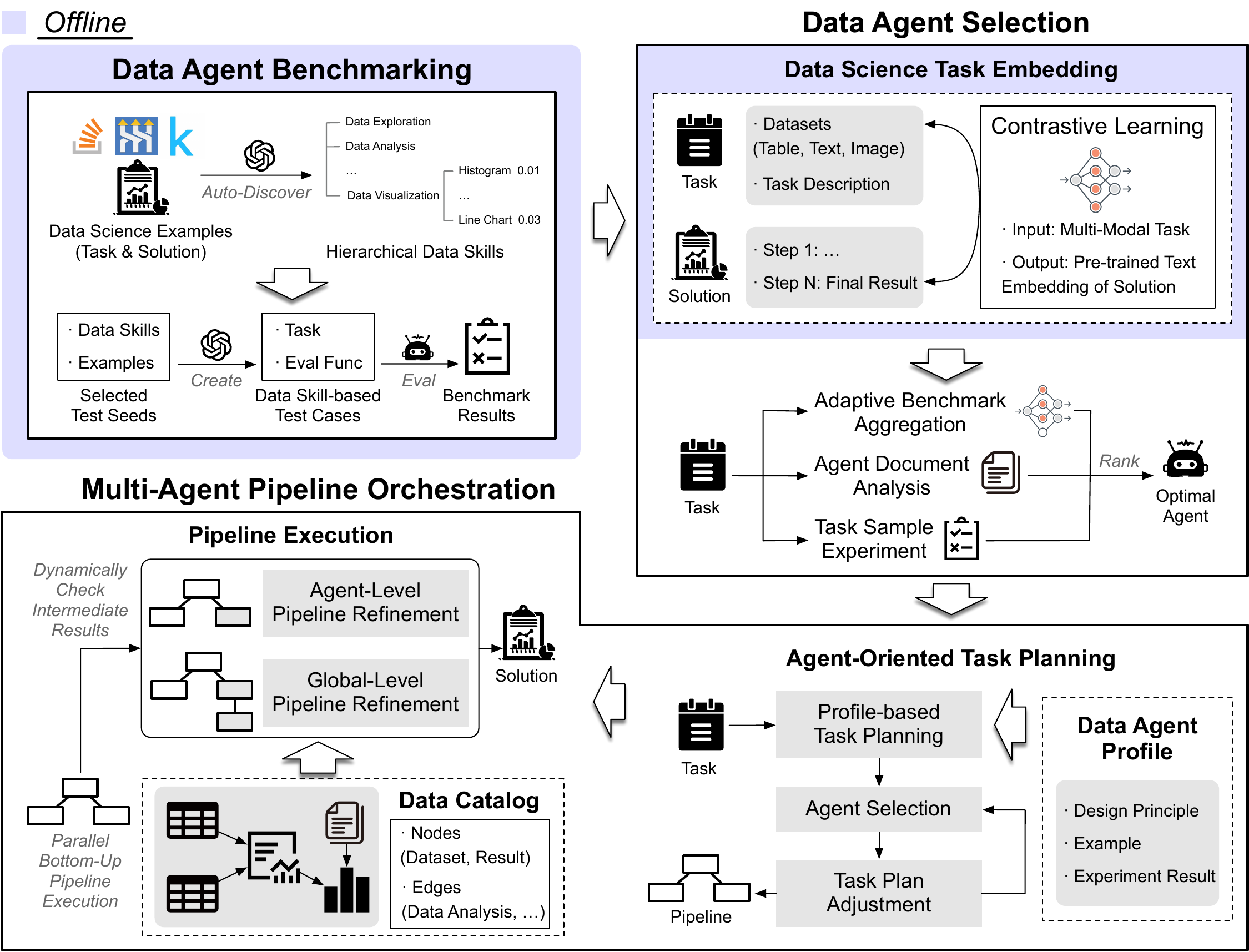}
\caption{Overview of \oursys. \label{fig:framework}}
\end{figure}

\section{\oursys: A Multi-Agent System on Data Science}
We first introduce the data agent architecture of \oursys (see Figure \ref{fig:framework}), and then present the components of \oursys.

\subsection{Overview of \oursys}

\oursys is designed to adaptively handle data science tasks by flexibly composing the complementary capabilities of diverse data agent, which is a challenging open research problem.
As shown in Figure \ref{fig:framework}, \oursys comprises an offline stage and an online stage.

\hi{Offline Data Agent Benchmarking.}
This stage aims to construct a comprehensive data agent benchmark that can cover diverse data science scenarios by composing basic data skills.
First, given a large corpus of data science examples, we employ \llm for quality filtering and data skill discovery.
Next, to organize the skills, we build a hierarchical structure via recursive clustering.
Each skill is also assigned an importance score based on its overall frequency or user-defined priorities.
Then, to reflect the capability requirements of specific data science scenario, we sample important skills probabilistically according to their scores and use \llm to generate corresponding test cases.

Besides, to ensure unbiased agent evaluation for an online task, benchmark test cases should be adaptively aggregated based on their similarity to the task.
We thus construct an efficient index to enhance the performance of similarity search over the test cases.
We will discuss the detailed techniques in Section \ref{sec:benchmark}.

\hi{Online Multi-Agent Pipeline Orchestration.}
Given an online data science task, this stage autonomously decomposes the task into a pipeline of sub-tasks aligned with data agent capabilities, selects an appropriate agent for each sub-task, and dynamically refines the pipeline to ensure both efficiency and robustness.

\emph{\textbf{(1) Data Agent Selection.}}
To effectively utilize benchmark results, we design a task embedding method to measure the similarity between benchmark test cases and the online task.
To this end, we fine-tune a task embedding model to align task embeddings with the pre-trained text embeddings of corresponding task solutions, which capture the capability requirements and reasoning complexity inherent in the tasks.
We then build an embedding index over the benchmark test cases.
For an online task, we utilize the index to efficiently retrieve top-$k$ relevant test cases, and evaluation scores of test cases are adaptively aggregated based on their similarity to guide agent selection.
The data agent with the highest aggregated score is selected as optimal.

Besides, for special cases where benchmark is unsuitable, we can also evaluate the agents through structured document analysis or via sample task experiments.
We will discuss the technical details in Section \ref{sec:select}.

\emph{\textbf{(2) Multi-Agent Pipeline Orchestration.}}
Given an online task, we first use \llms to decompose the task into a pipeline of interdependent sub-tasks using specialized agent profiles.
Each sub-task is assigned to an appropriate data agent selected as previously described.
We also iteratively refine the plan with additional adjustments such as sub-task merging or decomposition.

Then, we execute the pipeline in a parallel bottom-up manner based on its topological order.
To ensure robustness, we dynamically refine the pipeline based on intermediate results, including $(i)$ sub-task modification at agent level, and $(ii)$ global-level re-planning, where complete intermediate results are stored as datasets in our data catalog to prevent redundant computation.
Once all sub-tasks are executed, \oursys outputs the final task result to the user.
We will discuss the technical details in Section \ref{sec:orchetration}.

\hi{Integration of New Data Agent.}
\oursys is designed to be extensible, allowing integration of new data agents through agent profile construction based on agent document analysis.
Additionally, when sufficient time and resources are available, \oursys can further enhance the agent profile by executing the benchmark, thereby improving the accuracy of agent selection and pipeline orchestration.
Once integrated, the new agent can seamlessly collaborate with existing agents within our multi-agent pipeline orchestration framework.

\subsection{Data Agent Benchmarking}
\label{sec:benchmark}

Given the wide range of data science tasks across diverse application domains, existing benchmarks are typically constrained to a limited set of pre-defined task types~\cite{hu2024infiagent,jing2024dsbench}, and thus poorly evaluate data agents in more flexible use cases.
To address this limitation, we introduce a data skill-based benchmark, which can adaptively create test cases to cover different data science scenarios (see Figure \ref{fig:framework}).
Specifically, we first use \llms to extract data skills from a large corpus, whose compositions can represent the capability requirements of diverse tasks.
Next, we construct a hierarchical structure that captures the semantic relationships among these skills, with each skill assigned an importance score based on its overall frequency or user-defined priorities.
Then, during data agent evaluation, we can compose random sets of relevant skills tailored to the target scenario, and prompt \llms to generate corresponding test cases for adaptive and comprehensive assessment.

\subsubsection{Hierarchical Data Skill Discovery}
\label{sec:data-skill-discover}

\hi{Automatic Data Skill Discovery.}
To better understand the coverage of data science tasks, we first collect data science examples from diverse sources, including online websites (e.g., StackOverflow), professional competitions (e.g., Kaggle), and existing data science benchmarks.
Each data science example typically contains three components: $(i)$ ``task'': a natural language task description, as well as multi-modal datasets involved (e.g., tables, texts, images); $(ii)$ ``solution'': a detailed sequence of steps outlining the procedure of completing the task; $(iii)$ ``associated data skills'': the fundamental data science capabilities utilized in the solution, whose extraction procedure is described in the following paragraph.

We use \llms to automatically filter out high-quality examples and discover associated data skills in three steps. 
First, since the data science example can contain noisy content and intricate code snippets, we provide the task description and solution to \llms, and instruct it to summarize the procedure as a sequence of steps in concise natural language.
Second, we utilize \llms to evaluate the quality of example summaries, and exclude low-quality examples that either lack implementation details or provide only partial task solution.
Third, given the summarized steps of data science examples, we further use \llms to extract associated data skills~\cite{didolkar2024}.
For instance, given the example `Do lower-income students perform worse on a math test than higher-income students?', its solution reveals the data skills: `new column derivation', `filter by column', and `linear regression'.

\hi{Data Skill Hierarchy Construction.}
The mess of extracted data skills can be systematically organized into a hierarchical structure through recursively clustering semantically similar skills within the same group.
Specifically, we first use pre-trained text embeddings to represent the semantics of the data skills.
Then, these embeddings are clustered using Gaussian Mixture Models~\cite{sarthi2024raptor}, where Uniform Manifold Approximation and Projection method is used to reduce dimensionality by approximating the local data manifold~\cite{mcinnes2020umap}.
Next, for each skill cluster, we instruct \llm to generate a summary that captures the common skill theme of these grouped skills.
The cluster summaries serve as higher-level nodes in the hierarchy.
If the skill number within some cluster falls below a pre-defined threshold, such clusters are designated as leaf nodes.
Conversely, if some cluster still contains skills exceeding the threshold, the same clustering procedure is recursively applied to construct a more fine-grained hierarchical structure under the same parent cluster node.

\hi{Data Skill Weighting.}
To distinguish the relative importance of data skills, we initially use their occurrence frequencies across data science examples as a general proxy score.
Specifically, for each leaf node in the skill hierarchy, we compile the skills within the corresponding cluster, and count the number of examples in which any of these skills are associated.
The score of the leaf node is calculated as the ratio of its associated example count to the total number of examples.
Then, the scores of higher-level nodes can be recursively computed by aggregating the scores of their corresponding children nodes.

Besides, we also support users to adaptively adjust the scores of critical data skills for practical needs.
For instance, if users identify `missing value handling' as a more critical data cleaning skill, the score of corresponding node can be manually increased.
To maintain a constant total score among sibling nodes, the scores of the remaining sibling nodes are proportionally decreased.
The scores of descendant nodes are also adjusted accordingly, preserving their relative proportions within their respective sub-trees.

\begin{figure}[!t]
\centering
\includegraphics[width=.8\linewidth]{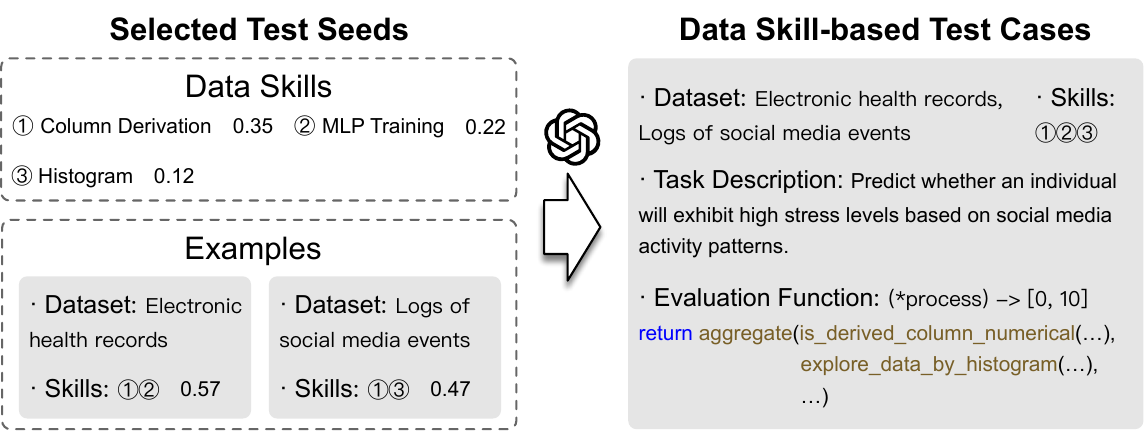}
\caption{Example of Data Skill-based Benchmark Construction. \label{fig:benchmark}}
\end{figure}

\subsubsection{Data Skill-based Benchmark Construction}
\label{sec:benchmark-construct}
The test case in the benchmark consists of two components: $(i)$ ``task'': a task description along with associated datasets; $(ii)$ ``evaluation function'': an executable function that outputs a scalar value within the range $[0, 10]$, quantifying the correctness and performance of data agents.
However, directly using \llms to generate test cases leads to low-quality benchmarks for two reasons. 
First, the generated test cases cannot fully capture the data science capabilities required for specific scenario.
Second, the test case may not adequately reflect conditions in real-world applications.

To address these challenges, we propose a data skill-based benchmark construction method (see Figure \ref{fig:benchmark}). 
First, we sample a random set of $k$ data skills from the leaf nodes of the hierarchy, with probabilities proportional to their importance scores.
Second, to enhance the realism of the generated test case, we retrieve data examples from the corpus that are pertinent to the selected skills, and incorporate them into \llm's input context for in-context learning.
Specifically, we assign a relevance score to each data science example based on its associated data skills.
The score of an example is computed as the sum of the importance scores of the $k$ selected skills present in its associated skill set. 
We select the examples with top scores as representative.

Then, we provide \llm with the selected data skills and data science examples, instructing it to synthesize test cases by leveraging both its pre-trained knowledge and in-context examples.
The synthesis process adheres to the following criteria:
$(i)$ the datasets must be drawn from those provided in the examples;
$(ii)$ the synthesized task and evaluation function should both involve the application of all the specified data skills; and $(iii)$ the evaluation function should be composed of a set of sub-functions, each of which returns a boolean value indicating whether a specific evaluation criterion is satisfied by the task results.
The final scalar score is then computed based on the aggregate results of these sub-functions, a design choice that promotes greater consistency across different test cases~\cite{liu2025reward}.

Two special cases also require consideration.
First, if the dataset size of some selected example exceeds the context window that \llms can effectively process~\cite{liu2024lost}, a subset of the dataset can be randomly sampled, supplemented with its metadata (e.g., column descriptions of the table) to preserve essential information.
Second, if certain evaluation standards are too ambiguous to be directly implemented as executable code, \llms can be invoked within the evaluation function to assess the task results guided by a set of generated rules.

Note that the complexity of test cases grows rapidly with the number of data skills $k$~\cite{yu2024skillmix}, which users can specify to tailor the benchmark to practical applications.

\subsection{Data Agent Selection}
\label{sec:select}
Since many data agents have been proposed to address similar data science tasks~\cite{hu2024infiagent,jing2024dsbench}, a straightforward approach to agent selection is to leverage their benchmark results. 
Specifically, we can assess agent performance by aggregating the evaluation scores of test cases and selecting the agent with the highest overall score.
However, this method overlooks the degree of relevance between benchmark test cases and the target task, and thus we should re-weight the test cases to ensure an unbiased assessment.

To address this issue, we first train an embedding model for data science tasks, where tasks with similar embeddings require similar data science competencies and reasoning process.
The unbiased evaluation score of an agent is then computed using these embeddings, as a weighted aggregation of the test case evaluation scores, where the weights reflect the similarity of each test case to the target task.
The agent achieving the highest score is regarded as the optimal choice.

Besides, we also propose alternative evaluation methods for the special cases where the benchmark is unsuitable.
For time-constrained settings, the agent is evaluated through structured document analysis using \llm, without actually executing the benchmark.
For tasks with extremely high execution time or resource demands (e.g., model training on large datasets), we generate samples from the target task and utilize them to further refine agent selection.

\begin{figure}[!t]
\centering
\includegraphics[width=\linewidth]{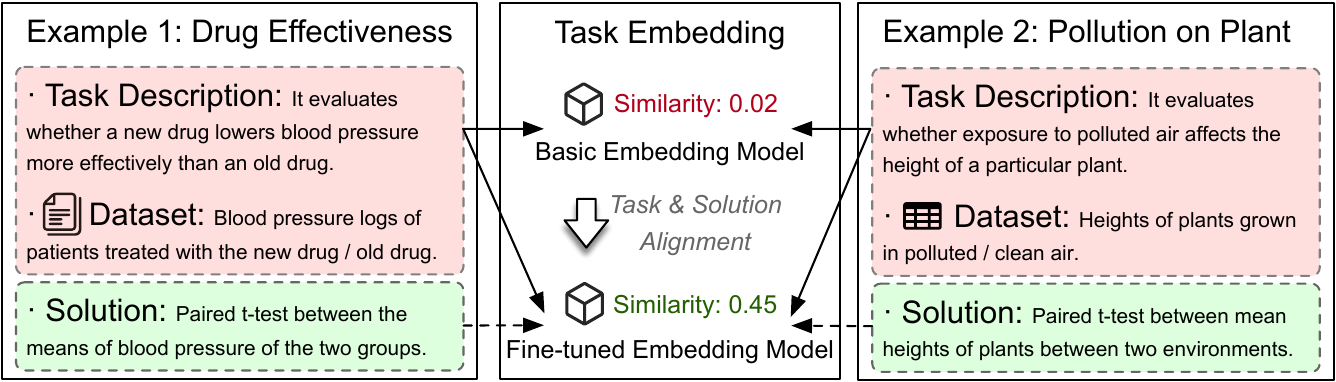}
\caption{Example of Fine-tuned Data Science Task Embedding. \label{fig:embed}}
\end{figure}

\subsubsection{Data Science Task Embedding.}
Since the data science task typically contains multi-modal datasets and textual descriptions, we can use multi-modal large language models (\mllms) to obtain its semantic embedding~\cite{jiang2025vlm2vec}, where both text and images are natively supported and tables can be represented as serialized text in CSV format.
Large datasets can be substituted with data samples and metadata to fit the context window of \mllm, as discussed in Section \ref{sec:benchmark-construct}.
However, such embedding method is sensitive to irrelevant information like $(i)$ domain of the task and $(ii)$ representation format of datasets, despite these factors minimally affect actual complexity of solving the task.

To overcome this limitation, we introduce a fine-tuning process that aligns data science task embedding with the pre-trained text embedding of its correct solution (see Figure \ref{fig:embed}).
We again use the corpus of high-quality data science examples annotated with associated data skills, as discussed in Section \ref{sec:data-skill-discover}. 
We employ a contrastive learning framework:  correct solutions to the task are treated as a positive example, while solutions to tasks involving mutually exclusive data skills serve as negative examples.
Using these training pairs, we compute the Multiple Negatives Ranking Loss~\cite{reimers2019sentence}, which minimizes embedding distances for positive pairs and maximizes embedding distances for negative pairs.
Once training converges, the embeddings exhibit meaningful alignment, allowing similarity metrics to reliably identify data science tasks with analogous solution procedures, which in turn reflect comparable performance of a data agent.

\begin{table}[!t]
\centering
\caption{Comparison of Data Agent Selection Methods.}
\label{tab:agent-select}
\begin{tabular}{|c|c|c|c|}
\hline
\textbf{Method}  & \textbf{Accuracy} & \textbf{Online Overhead} & \textbf{Offline Overhead} \\ \hline
Adaptive Benchmark Aggregation & Mid & Low & High \\ \hline
Agent Document Analysis & Low & Mid & Low \\ \hline
Task Sample Experiment & High & High & Low \\ \hline
\end{tabular}
\end{table}

\subsubsection{Heterogeneous Data Agent Selection}
\label{sec:heterogeneous-select}
We propose three methods for selecting the most suitable data agent for a given task, i.e., adaptive benchmark aggregation, agent document analysis, and task sample experiment, each offering distinct advantages and disadvantages (see Table \ref{tab:agent-select}).

\hi{Adaptive Benchmark Aggregation.}
In the offline stage, we run the benchmark using the candidate data agents, resulting in each agent associated with evaluation scores corresponding to each test case.
We also build an embedding index over the test cases using fine-tuned task embedding model to support efficient similarity search.
Then, for an online task, we similarly generate its embedding and identify the most relevant test cases with top-$k$ embedding similarities.
The weights assigned to these $k$ selected test cases are their normalized similarities such that their sum is equal to one.
We thus obtain an unbiased evaluation score for each agent, which is computed by the weighted sum of its evaluation scores across these selected test cases.
The agents are then ranked by the aggregated scores, and the top one with the highest score is identified as the most suitable choice.

\hi{Agent Document Analysis.}
Sometimes, it is impractical to evaluate the data agent by executing time-intensive benchmark.
For instance, when integrating a new agent into \oursys, a temporary evaluation method is required to enable its inclusion in the agent selection framework prior to completing its full benchmark evaluation.
Thus, given an online task, we can assess the data agent by using \llm to directly analyze agent documents (e.g., research paper, repository Readme files).

Specifically, we instruct \llms to focus on three structured artifacts of the document, which are critical for agent evaluation: 
$(i)$ Design Principle.
It describes the intended usage scenarios and technical contributions of the agent, which are often extracted from the introductory sections of the documents.
We can use \llms to compare the given task with these design specifications, determining whether the agent is appropriately targeted for the task.
$(ii)$ Representative Example. 
It can be treated analogously to test cases in the benchmark, where higher embedding similarity to the given task indicates greater suitability of the agent.
$(iii)$ Experiment Result. 
It describes the experimental results of the agent on various datasets in comparison with baselines.
First, if details regarding the datasets or baselines are missing, we can prompt \llms to use search engine as tools to retrieve and complete the missing information. Then, we use \llms to compare the target task with the experimental datasets, and predict the agent’s performance based on reported experimental results.

Finally, guided by the analysis of aforementioned three key artifacts, we employ \llms to generate an evaluation score in the range $[0, 10]$, which is comparable to benchmark scores.

\hi{Task Sample Experiment.}
If the given task involves large datasets or demands large resources, agent selection can be further refined through sample experiments. 
Specifically, we randomly sample the task dataset to construct a task sample. Next, we select the agents with the top scores based on the evaluation and execute the task sample using these candidates. Finally, we compare their execution results and use \llms to determine which agent most effectively solves the task sample.

\subsection{Multi-Agent Pipeline Orchestration}
\label{sec:orchetration}
For complex data science tasks that exceed the capabilities of any single data agent~\cite{google2025,hong2024data,you2025datawiseagent}, it is essential to compensate for their limitations through a unified collaborative framework. 
However, most existing work define a deterministic pipeline with a fixed set of data agents based on human expertise, limiting their ability to leverage the broader range of available agents through extensible integration~\cite{AutoML_Agent}.
To this end, we propose a flexible and adaptive pipeline orchestration algorithm that accommodates diverse tasks and agents, and effectively manages the intricate reasoning process through agent collaboration.
Specifically, we first introduce an agent-oriented task planning approach that decomposes the given task into a pipeline of sub-tasks by leveraging specialized agent profiles, heterogeneous agent selection for each sub-task, and \llm-based plan adjustment to ensure both effectiveness and efficiency.
Next, we execute the pipeline interactively, with dynamic pipeline refinement to adapt to intermediate results through either agent-level modifications or holistic re-planning.

\subsubsection{Agent-Oriented
Task Planning}
\label{sec:task-plan}
Following existing studies~\cite{li2025agentplanning}, we feed the given task and a set of agent profiles to \llms, and instruct \llms to generate a task plan comprising sub-tasks, each associated with a relevant subset of datasets and aligned with capabilities of some agents. We further prompt \llms to evaluate the task plan for completeness and non-redundancy, ensuring that the sub-tasks collectively address the given task in its entirety while avoiding unnecessary or duplicate sub-tasks. Our approach is distinguished by three key aspects: $(i)$ design of specialized data agent profiles, $(ii)$ heterogeneous agent selection for each sub-task, and $(iii)$ adaptive task plan adjustment.

\hi{Data Agent Profile Construction.}
To fully describe the capabilities of the data agent, we construct the agent profile comprising three components: $(i)$ ``design principle'', $(ii)$ ``representative example'', and $(iii)$ ``experiment result'', paralleling the key artifacts extracted from agent document as discussed in Section~\ref{sec:heterogeneous-select}. We can enhance the agent profile by including its benchmark results in addition to document information. Specifically, for the “representative example” component of an agent, we incorporate benchmark test cases by quantifying its performance deviation on each test case $t_i$ as $\frac{s_{i} - mean_i}{mean_i}$, where $s_{i}$ denotes the evaluation score of the given agent on $t_i$, and $mean_i$ represents the average score of $t_i$ across all data agents. Test cases exhibiting the highest and lowest deviation scores are selected as positive and negative examples respectively.
For the ``experiment result'' component, we further incorporate overall benchmark metrics (e.g., accuracy, recall), using the seed data skills associated with the benchmark as descriptive metadata. Note that our agent profile design is also compatible with the ``Agent Card'' interface defined by the Agent-to-Agent (A2A) protocol~\cite{google2025a2a}.

\hi{Heterogeneous Data Agent Selection.}
To construct the context for each sub-task during agent selection, we first employ \llms to analyze the dependency relationship among sub-tasks in the task plan, representing them as a directed graph with sub-tasks as nodes and dependencies as edges. Next, for a given sub-task, we treat the intermediate results of its dependent sub-tasks as auxiliary datasets, and prompt \llms to synthesize mock dataset descriptions to serve as the sub-task input. Then, we apply the agent selection method described in Section \ref{sec:select} to identify the most suitable agent for the sub-task.

\hi{Task Plan Adjustment.}
After sub-task decomposition and agent assignment, we further refine the task plan by applying adjustments such as additional decomposition and merging. First, if no suitable data agent can be selected for a given sub-task (e.g., all agents with evaluation scores below a pre-defined threshold), we instruct \llm to revise the task plan by further decomposing the sub-task into simpler sub-tasks.

Besides, to enhance efficiency of task plan, we attempt to merge correlated sub-tasks, thereby reducing the complexity of execution pipeline.
Specifically, we first provide the dependency graph of sub-tasks to \llms and instruct \llms to identify sub-graphs where the nodes $(i)$ involve interrelated datasets, and $(ii)$ can be coherently aggregated into a larger, logically consistent sub-task. Next, for each identified sub-graph, we prompt \llms to generate a consolidated sub-task that summarizes the constituent sub-tasks within the sub-graph. Then, we apply the heterogeneous data agent selection method to identify a suitable agent for the new sub-task.  If a suitable agent is successfully selected, the original sub-graph is replaced with the new sub-task and its corresponding agent assignment.

\subsubsection{Pipeline Execution}
Unlike traditional approaches that depend on fixed data agents and rigid workflows, \oursys is designed to dynamically adapt to the diversity and unpredictability of both data agents and data science tasks. As illustrated in Figure \ref{fig:framework}, \oursys can execute its pipeline interactively, dynamically refining its pipeline at both the agent level and the global level based on intermediate results and execution states.

\hi{Parallel Pipeline Execution.}
To enhance execution efficiency, following the dependency graph of sub-tasks, we execute the sub-tasks in parallel bottom-up along their topological order in the graph. Once their dependent sub-tasks are completed, each sub-task is executed using the corresponding intermediate results, proceeding iteratively until all sub-tasks are completed and the final result is obtained. During execution, each sub-task is carried out by its assigned agent, after which we use \llms to verify whether the generated intermediate results adequately fulfill the sub-task. If discrepancies are identified, they trigger dynamic refinement of the pipeline to ensure robustness, as described next.

\hi{Agent-Level Pipeline Refinement.}
When a sub-task fails to be resolved by the assigned data agent, we first use \llms to reflect potential failure causes based on the sub-task input, including: $(i)$ sub-task description, $(ii)$ dataset, and $(iii)$ intermediate results. For example, if a failure stems from an ambiguous sub-task description that the agent misinterprets, we use \llms to rephrase it, drawing on insights from the failure logs. Next, if required datasets or intermediate results are missing, we use \llms to further examine the available datasets and sub-tasks to identify supplementary information.
Then, if intermediate results are improperly formatted, \llms are employed to refine it by further analyzing the solution process of the corresponding sub-task.

Besides, if a failure cannot be attributed to unexpected sub-task inputs, it may result from an incorrect agent selection. In such cases, we review the ranking of data agents based on their evaluation scores for the given sub-task, and select the next-best agent with a slightly lower score than the previously chosen one.

Lastly, we re-execute the sub-task using refined input information or a newly assigned data agent.

\hi{Global-Level Pipeline Refinement.}
If a sub-task failure cannot be resolved via agent-level refinement, a holistic re-planning of the entire pipeline is required.
First, during pipeline execution, we systematically store the intermediate results of sub-tasks in a data catalog (see Figure~\ref{fig:framework}), which ensures preservation of previously computed results and avoids redundant computations.  Specifically, the output generated by each sub-task is treated as a new dataset. For this dataset, metadata is created by \llms based on the sub-task's input and the process used to solve it. These newly formed datasets are then incorporated into the data catalog. Then, during task re-planning, we apply the agent-oriented task planning algorithm in Section~\ref{sec:task-plan}, treating all elements in the data catalog as available datasets.

\section{Data Analytics Agents}
\label{sec:Analytics}

We first summarize the overview of data analytics agents~\cite{wang2025aop,iDataLake}. We then introduce four data analytics agents, including unstructured data analytics agents, semantic structured data analytics agents, data lake analytics agents, and multi-modal data analytics agents.

\hi{Overview of Data Analytics Agent.}  In the offline phase, the data analytics agent generates a semantic catalog and builds semantic indexes for a variety of data types. It also defines semantic operators, such as semantic filter, semantic group-by, semantic sorting, semantic projection, semantic join, and others. Each semantic operator is represented logically (e.g., as an entity that satisfies certain conditions) to facilitate matching natural language segments with these semantic operators. Each semantic operator is also associate with multiple physical operators, e.g., execution by \llms, pre-programmed functions, \llm coding, etc. When processing an NL query, the data analytics agent decomposes the query into sub-tasks and orchestrates them into a pipeline. The agent then optimizes this pipeline by selecting the optimal sequence of semantic operators and executes the pipeline efficiently by calling the semantic operators.

\hi{Unstructured Data Analytics Agent.} It supports semantic analytics on unstructured data using natural language queries~\cite{wang2025aop}. The challenges include orchestrating a natural language query into a pipeline, self-reflecting on the pipeline, optimizing the pipeline for low cost and high accuracy, and executing the pipeline efficiently. We propose a logical plan generation algorithm that constructs logical plans capable of solving complex queries through correct logical reasoning. Additionally, we introduce physical plan optimization techniques that transform logical plans into efficient physical plans, based on a novel cost model and semantic cardinality estimation. Finally, we design an adaptive execution algorithm that dynamically adjusts the plan during execution to ensure robustness and efficiency. In addition, we can extract a semantic catalog for unstructured data and use it to guide pipeline orchestration.

\hi{Semantic Structured Data Analytics Agent.} Existing database systems operate under a closed-world model, which limits their ability to support open-world queries. To overcome these limitations, we integrate databases with LLMs to enhance the capabilities of database systems. By using LLMs as semantic operators, we can support open-world data processing functions such as semantic acquisition, extraction, filtering, and projection. This approach allows us to extend SQL to incorporate these LLM-powered semantic operators, creating what we call semantic SQL. Additionally, NL queries can be transformed into semantic SQL using specialized NL2SQL agents. To execute semantic SQL effectively, we propose three techniques. First, we replace some semantic operators with traditional operators. For instance, instead of using ``semantic acquire the capital of China,'' we can use `city = Beijing.' Second, we design a multi-step filtering process to accelerate the processing of semantic operators, including embedding-based filtering and small LLM-based filtering. Third, we estimate the cost of semantic operators to determine the most efficient order for executing multiple semantic operations.

\hi{Data Lake Analytics Agent.}  Its aim is to perform data analytics on semi-structured and unstructured datasets~\cite{iDataLake}. However, integrating \llms into data analytics workflows for data lakes remains an open research problem due to the following challenges: heterogeneous data modeling and linking, semantic data processing, automatic pipeline orchestration, and efficient pipeline execution. To address these challenges, we propose a data lake agent designed to handle data analytics queries over data lakes. We introduce a unified embedding approach to efficiently link heterogeneous data types within a data lake. Additionally, we present a set of semantic operators tailored for data analytics over data lakes. Our iterative two-stage algorithm facilitates automatic pipeline orchestration, incorporating dynamic pipeline adjustment during query execution to adapt to intermediate results. Overall, the data lake agent represents a significant advancement in enabling high-accuracy, practical data analytics on data lakes. Unlike previous approaches that rely on lossy data extraction or are constrained by SQL's rigid schema, the data lake agent effectively employs the semantic understanding capabilities of LLMs to provide a more comprehensive and efficient solution.

\hi{Multi-modal Data Analytics Agent.} We also design data agents to support multi-modal data, e.g., audio, video. First, the integration and management of heterogeneous data types, such as text, images, audio, and video, are critical yet challenging tasks that require robust frameworks for seamless merging into cohesive datasets. Representing the multi-modal data in a unified format demands advanced embedding techniques to maintain unique characteristics while enabling analysis. Semantic understanding across different modalities is essential for extracting meaningful insights, necessitating sophisticated NLP, computer vision, and audio processing algorithms. Additionally, designing a flexible query system capable of interpreting and executing complex multi-modal queries is crucial. Aligning and fusing the data from various modalities ensures a coherent data view, while scalable methods are necessary for handling increasing data volumes efficiently.

\section{DBA Agent}
\label{sec:DBA}

Database administrators (DBAs) often face challenges managing multiple databases while providing prompt responses, as delays of even a few hours can be unacceptable in many online scenarios. Current empirical methods offer limited support for database diagnosing issues, further complicating this task. To address these challenges, we propose a DBA agent, which is a database diagnosis system powered by LLMs~\cite{zhou2024dbot,sun2025dbot}. This system autonomously acquires knowledge from diagnostic documents and generates well-founded reports to identify root causes of database anomalies accurately. The DBA agent includes several key components. The first is to extract knowledge from documentation automatically. The second is to generate prompts based on knowledge matching and tool retrieval. The third conducts root cause analysis using a tree search algorithm. The fourth optimizes  execution pipelines for high efficiency. Our results demonstrate that the DBA agent significantly outperforms traditional methods and standard models like GPT-4~\cite{openai2024gpt4o} in analyzing previously unseen database anomalies. 

\section{Opportunities and Challenges}

\hi{Theoretical Guarantee.} A Data Agent may not always deliver 100\% accurate results due to potential hallucinations from \llms and semantic operators~\cite{huang2025hallucination}. Therefore, it is crucial to provide a theoretical guarantee for the reliability of data agent systems.

\hi{Self-Reflection and Reward Model.} A Data Agent needs to continuously enhance its accuracy and efficiency. Providing feedback to the Data Agent is essential for its self-improvement, and designing effective self-reflection techniques and reward models are viable strategies to tackle these challenges.

\hi{Data Agent Benchmark.} It is crucial to develop benchmarks for Data Agents to evaluate these systems effectively, particularly in areas like data science, data analytics, and database administration. 

\hi{Security and Privacy.} Ensuring security and privacy in data agents involves protecting sensitive information from unauthorized access while maintaining compliance with privacy regulations.

\hi{Scalability and Performance.} Scalability and performance challenges for data agents involve efficiently managing large and complex datasets while maintaining high processing speed and accuracy. As data volume and complexity grow, agents must be designed to scale seamlessly without degradation in performance.

\section{Conclusion}

In this paper, we propose the concept of a Data Agent for autonomously supporting Data+AI applications. We design a comprehensive architecture for developing a Data Agent, which includes components for data understanding and exploration, data engine comprehension and scheduling, and pipeline orchestration. We also present examples such as the data science agent, data analytics agents, and a DBA agent. The introduction of the Data Agent will present numerous challenges that require attention from the data community to address effectively. Moreover, there are numerous opportunities to develop data agents in various areas, including database development, database design, data transformation, data flywheel, and data fabric, etc.

\small
\section*{Acknowledgments}
This paper was supported by National Key R\&D Program of China (2023YFB4503600),
NSF of China (62525202, 62232009), Shenzhen Project (CJGJZD20230724093403007), Zhongguancun
Lab, Huawei, and Beijing National Research Center for Information Science and Technology
(BNRist). Guoliang Li is the corresponding author.


\begin{thebibliography}{10} 
\itemsep=1pt 


\bibitem{li2021ai} G. Li, X. Zhou, L. Cao.  \newblock AI meets database: AI4DB and DB4AI. \newblock \emph{SIGMOD}, 2021.

\bibitem{wu2024automatic} Y. Wu, X. Zhou, Y. Zhang, G. Li.  \newblock Automatic Database Index Tuning: A Survey. \newblock \emph{TKDE}, 2024.

\bibitem{zhou2024breaking} W. Zhou, C. Lin, X. Zhou, G. Li.  \newblock Breaking It Down: An In-Depth Study of Index Advisors. \newblock \emph{VLDB}, 2024.

\bibitem{han2023dynamic} Y. Han, C. Chai, J. Liu, G. Li, C. Wei, et al.  \newblock Dynamic materialized view management using graph neural network. \newblock \emph{ICDE}, 2023.

\bibitem{zhou2023grep} X. Zhou, G. Li, J. Feng, L. Liu, W. Guo.  \newblock Grep: A graph learning based database partitioning system. \newblock \emph{SIGMOD}, 2023.

\bibitem{zhao2023automatic} X. Zhao, X. Zhou, G. Li.  \newblock Automatic database knob tuning: A survey. \newblock \emph{TKDE}, 2023.

\bibitem{marcus2021bao} R. Marcus, P. Negi, H. Mao, N. Tatbul, M. Alizadeh, et al.  \newblock Bao: Making learned query optimization practical. \newblock \emph{SIGMOD}, 2021.

\bibitem{zhou2021mcts} X. Zhou, G. Li, C. Chai, J. Feng.  \newblock A learned query rewrite system using monte carlo tree search. \newblock \emph{VLDB}, 2021.

\bibitem{yu2022hybrid} X. Yu, C. Chai, G. Li, J. Liu.  \newblock Cost-based or learning-based? A hybrid query optimizer for query plan selection. \newblock \emph{VLDB}, 2022.

\bibitem{wang2021face} J. Wang, C. Chai, J. Liu, G. Li.  \newblock FACE: A normalizing flow based cardinality estimator. \newblock \emph{VLDB}, 2021.

\bibitem{sun2021exploration} J. Sun, J. Zhang, Z. Sun, G. Li, N. Tang. \newblock Learned cardinality estimation: A design space exploration and a comparative evaluation. \newblock \emph{VLDB}, 2021.

\bibitem{hilprecht2022zeroshot} B. Hilprecht and C. Binnig. \newblock Zero-Shot Cost Models for
Out-of-the-box Learned Cost Prediction. \newblock \emph{VLDB}, 2022.

\bibitem{wehrstein2025foundation} J. Wehrstein, C. Binnig, F. Özcan, S. Vasudevan, Y. Gan, et al. \newblock Towards Foundation Database Models. \newblock \emph{CIDR}, 2025.

\bibitem{zhou2020predict} X. Zhou, J. Sun, G. Li, J. Feng.  \newblock Query performance prediction for concurrent queries using graph embedding. \newblock \emph{VLDB}, 2020.

\bibitem{sun2023index} Z. Sun, X. Zhou, G. Li.  \newblock Learned index: A comprehensive experimental evaluation. \newblock \emph{VLDB}, 2023.

\bibitem{guo2025indatabase} Y. Guo, G. Li, R. Hu, Y. Wang.  \newblock In-database query optimization on SQL with ML predicates. \newblock \emph{The VLDB Journal}, 34 (1), 12, 2025.

\bibitem{chai2023goodcore} C. Chai, J. Liu, N. Tang, J. Fan, D. Miao, et al.  \newblock In-database query optimization on SQL with ML predicates. \newblock \emph{SIGMOD}, 2023.

\bibitem{siddiqi2023saga} S. Siddiqi, R. Kern, M. Boehm.  \newblock SAGA: A Scalable Framework for Optimizing Data Cleaning Pipelines for Machine Learning Applications. \newblock \emph{SIGMOD}, 2023.

\bibitem{wang2025aop} J. Wang, G. Li.  \newblock Aop: Automated and interactive llm pipeline orchestration for answering complex queries. \newblock \emph{CIDR}, 2025.

\bibitem{zhou2023febench} X. Zhou, C. Chen, K. Li, B. He, M. Lu, et al.  \newblock Febench: A benchmark for real-time relational data feature extraction. \newblock \emph{VLDB}, 2023.

\bibitem{moritz2018ray} P. Moritz, R. Nishihara, S. Wang, A. Tumanov, R. Liaw, et al.  \newblock Ray: A Distributed Framework for Emerging AI Applications. \newblock \emph{OSDI}, 2018.

\bibitem{openai2024gpt4o} A. Hurst, A. Lerer, A. P. Goucher, A. Perelman, A. Ramesh, et al.  \newblock GPT-4o System Card. \newblock \emph{arXiv:2410.21276}, 2024.

\bibitem{guo2025deepseek} D. Guo, D. Yang, H. Zhang, J. Song, R. Zhang, et al.  \newblock Deepseek-r1: Incentivizing reasoning capability in llms via reinforcement learning. \newblock \emph{arXiv:2501.12948}, 2025.


\bibitem{anthropic2024mcp} Anthropic. \newblock Introducing the Model Context Protocol. \newblock \emph{Newsroom \ Anthropic}, 2024.

\bibitem{google2025a2a} R. Surapaneni, M. Jha, M. Vakoc, T. Segal. \newblock Announcing the Agent2Agent Protocol (A2A). \newblock \emph{Google for Developers}, 2025.


\bibitem{hu2024infiagent} X. Hu, Z. Zhao, S. Wei, Z. Chai, Q. Ma, et al.  \newblock InfiAgent-DABench: evaluating agents on data analysis tasks. \newblock \emph{ICML}, 2024.

\bibitem{jing2024dsbench} L. Jing, Z. Huang, X. Wang, W. Yao, W. Yu, et al.  \newblock DSBench: How Far Are Data Science Agents to Becoming Data Science Experts?. \newblock \emph{ICLR}, 2025.

\bibitem{didolkar2024} A. Didolkar, A. Goyal, N. R. Ke, S. Guo, M. Valko, T. Lillicrap, et al.  \newblock Metacognitive capabilities of llms: An exploration in mathematical problem solving. \newblock \emph{NIPS}, 2024.

\bibitem{sarthi2024raptor} P. Sarthi, S. Abdullah, A. Tuli, S. Khanna, A. Goldie, et al.  \newblock RAPTOR: Recursive Abstractive Processing for
Tree-Organized Retrieval. \newblock \emph{ICLR}, 2024.

\bibitem{mcinnes2020umap} L. McInnes, J. Healy, J. Melville.  \newblock UMAP: Uniform Manifold Approximation and Projection for Dimension Reduction. \newblock \emph{arXiv:1802.03426}, 2020.

\bibitem{liu2025reward} Z. Liu, P. Wang, R. Xu, S. Ma, C. Ruan, et al.  \newblock Inference-time scaling for generalist reward modeling. \newblock \emph{arXiv:2504.02495}, 2025.

\bibitem{liu2024lost} N. F. Liu, K. Lin, J. Hewitt, A. Paranjape, M. Bevilacqua, et al.  \newblock Lost in the Middle: How Language Models Use Long Contexts. \newblock \emph{Transactions of the Association for Computational Linguistics}, 12, 2024.

\bibitem{yu2024skillmix} D. Yu, S. Kaur, A. Gupta, J. Brown-Cohen, A. Goyal, et al.  \newblock Skill-Mix: a Flexible and Expandable Family of Evaluations for AI models. \newblock \emph{ICLR}, 2024.


\bibitem{jiang2025vlm2vec} Z. Jiang, R. Meng, X. Yang, S. Yavuz, Y. Zhou, et al. \newblock Vlm2vec: Training vision-language models for massive multimodal embedding tasks. \newblock \emph{ICLR}, 2025.

\bibitem{reimers2019sentence} N. Reimers, I. Gurevych. \newblock Sentence-{BERT}: Sentence Embeddings using {S}iamese {BERT}-Networks. \newblock \emph{EMNLP-IJCNLP}, 2019.


\bibitem{google2025} J. Fine, M. Kolla, I. Soloducho. \newblock Data Science Agent in Colab: The future of data analysis with Gemini. \newblock \emph{Google for Developers}, 2025.

\bibitem{hong2024data} S. Hong, Y. Lin, B. Liu., B. Liu, B. Wu, et al.  \newblock Data Interpreter: An LLM Agent For Data Science. \newblock \emph{arXiv:2402.18679}, 2024.

\bibitem{you2025datawiseagent} Z. You, Y. Zhang, D. Xu, Y. Lou, Y. Yan, et al.  \newblock DatawiseAgent: A Notebook-Centric LLM Agent Framework for Automated Data Science. \newblock \emph{arXiv:2503.07044}, 2025.

\bibitem{AutoML_Agent} P. Trirat, W. Jeong, Wonyong, S. J. Hwang.  \newblock Auto{ML}-Agent: A Multi-Agent {LLM} Framework for Full-Pipeline Auto{ML}. \newblock \emph{ICML}, 2025.

\bibitem{li2025agentplanning} A. Li,Y. Xie, S. Li,F. Tsung, B. Ding, et al. \newblock Agent-Oriented Planning in Multi-Agent Systems. \newblock \emph{ICLR}, 2025.

\bibitem{iDataLake} J. Wang, G. Li, J. Feng. \newblock iDataLake: An LLM-Powered Analytics System on Data Lakes. \newblock \emph{IEEE Data Eng. Bull.}, 49(1), 57-69, 2025.

\bibitem{zhou2024dbot} X. Zhou, G. Li, Z. Sun, Z. Liu, W. Chen, et al. \newblock D-bot: Database diagnosis system using large language models. \newblock \emph{VLDB}, 2024.

\bibitem{sun2025dbot} Z. Sun, X. Zhou, J. Wu, W. Zhou, G. Li. \newblock D-Bot: An LLM-Powered DBA Copilot. \newblock \emph{SIGMOD-Companion}, 2025.


\bibitem{huang2025hallucination} L. Huang, W. Yu, W. Ma, W. Zhong, Z. Feng, et al. \newblock A survey on hallucination in large language models: Principles, taxonomy, challenges, and open questions. \newblock \emph{ACM Transactions on Information Systems}, 43(2), 1-55, 2025.

\end{thebibliography}
\end{document}